# Low-Speed ADC Sampling Based High-Resolution Compressive Channel Estimation

Guan Gui
Dept. of Communication Engineering
Tohoku University
Sendai, Japan
gui@mobile.ecei.tohoku.ac.jp

Aihua Kuang
Dept of Electronics
School of Electronics and Information
Engineering of Zhengzhou, China
yi_1978@163.com

Ling Wang
Dept. of Electronic Engineering
UESTC
Chengdu, China
eewangling@gmail.com

*Abstract*—Broadband channel is often characterized by a sparse multipath channel where dominant multipath taps are widely separated in time, thereby resulting in a large delay spread. Traditionally, accurate channel estimation is done by sampling received signal by analog-to-digital converter (ADC) at Nyquist rate (high-speed ADC sampling) and then estimate all channel taps with high-resolution. However, traditional linear estimation methods have two mainly disadvantages: 1) demand of the high-speed ADC sampling rate which already exceeds the capability of current ADC and also the high-speed ADC is very expensive for regular wireless communications; 2) neglect the inherent channel sparsity and the low spectral efficiency wireless communication is unavoidable. To solve these challenges, in this paper, we propose a high-resolution compressive channel estimation method by using low-speed ADC sampling. Our proposed method can achieve close performance comparing with traditional sparse channel estimation methods. At the same time, the proposed method has following advantages: 1) reduce communication cost by utilizing cheap low-speed ADC; 2) improve spectral efficiency by extracting potential training signal resource. Numerical simulations confirm our proposed method using low-speed ADC sampling.

## I. INTRODUCTION

With the number of wireless subscribers increasing every day, various wireless devices, e.g., smart phones, computers and laptops, generate massive data traffic on the rise. An authoritative industry report predicts that mobile generated traffic will exceed that from fixed personal computers (PCs) by 2015, underscoring the fact that most information and communication technology services may be expected to migrate to portable mobile devices over the next a few years. This trend toward mobile platforms has significant implications for radio technology and wireless networks, which will enable this paradigm shift, as well as for the wide variety of internet applications currently supported by fixed network devices such as PCs and televisions [1]. To satisfy the greedy demand of high-speed data services, high data rate broadband communication is an indispensable technique [2] in the next-generation communication systems.

However, broadband signal transmission over multipath channel is often susceptible to frequency-selective fading. In the sequel, the broadband channel is described by sparse channel model which multipath taps are widely separated in time, thereby creating a large delay spread [3]. According to Nyquist sampling theorem, the sampled channel length is increasing inevitably due to broadband nature of communication channels, and the length of a sampled channel easily reaches hundreds of taps [4]. A typical example of sparse channel is shown in Fig. 1. The well-known approach becomes impractical when the bandwidth is too large because it is challenging to build sampling hardware that operates at a sufficient sampling rate. The demands of many modern applications exceed the capabilities of current technology. Even though recent developments in analog-to-digital converter (ADC) technologies have increased the sampling speed, state-of-the-art architectures are not yet adequate for high-dimensional signal processing [5]. Except of the incapability, high-speed ADC is very expensive in general and cannot be utilized widely. Hence, it is necessary to develop novel alternative technique. To relax the strict requirement of high speed ADC sampling, recently, some pioneering works have been done in [5-7]. In these works, different ADCs working at sub-Nyquist sampling rate have been proposed. However, these pioneering works focus on the theoretical analysis of using compressive sensing (CS) [8,9]. All of the works have not considered their applications on channel estimation in broadband wireless communication systems.

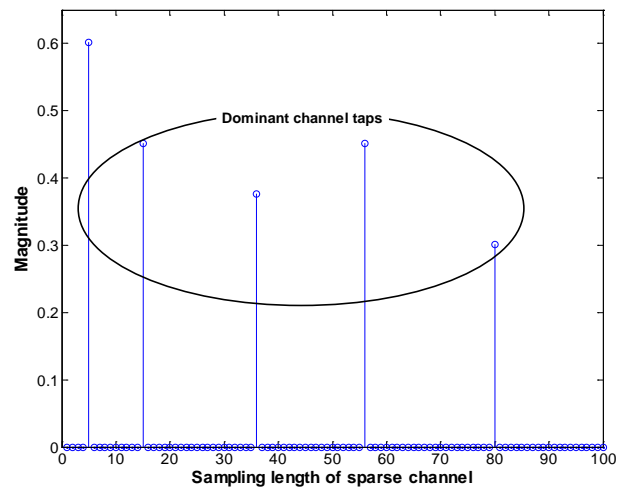

Fig.1. A typical example of sparse multipath channel where the overall sampling length is 100 while the number of dominant channel taps is 5 and most of channel taps are zeros or close to noise level.

In the traditional sparse multipath communication system, the receiver is often equipped with high speed ADC which is shown in Fig. 2. High-resolution sparse channel estimation method using compressive sampling matching pursuit algorithm (CoSaMP) [10] has been proposed in our previous work [11]. However, the high speed ADC costs much more for the ordinary wireless communications. In addition, increasing transmission bandwidth requires much higher ADC sampling rate at the receiver. This well-known approach becomes impractical since it is difficult to build sampling hardware that operates at a sufficient Nyquist rate sampling. Equipping low speed ADC at the receiver is a good candidate to solve the challenge. However, low-speed ADC sampling will result in low-resolution channel estimation with traditional estimation methods. The worst case is that low-resolution channel estimation often deteriorates equalization at the receiver. To solve the contradiction between high-resolution channel estimation and low communication costs, it is necessary to develop high-resolution compressive channel estimation techniques. In this paper, different from the traditional method, we assume that the receiver is equipped with parallel low-speed ADCs as shown in Fig. 3. Based on the low-speed ADC sampling system model, we propose a high-resolution compressive channel estimation method better than the conventional method in our previous work [11]. Numerical results also confirm the advantage of our proposed method.

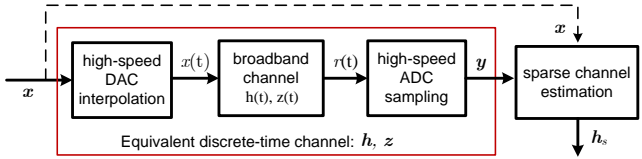
Fig. 2. Broadband system model based on the high-speed ADC sampling.

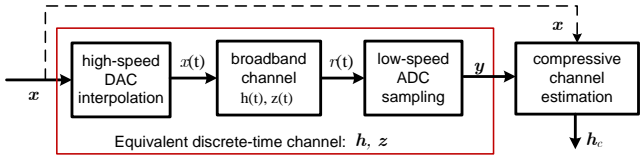
Fig. 3. Broadband system model based on the low-speed ADC sampling.

Section II introduces the system model and problem formulation. Section III discusses compressive channel estimation for broadband communication systems under low-speed ADC sampling. In section IV various numerical simulation results and discuss on their performance comparison are given. Concluding remarks are presented in Section V.

Notations: In this paper, we use boldface lower case letters $\boldsymbol{x}$ to denote vectors, boldface capital letters $\boldsymbol{X}$ to denote matrices. x represents the complex Gaussian random variable. E[.] stands for the expectation operation and $\boldsymbol{X}$, $\boldsymbol{X}^\dagger$ denote the matrix $\boldsymbol{X}^H$ transposition and conjugated transposition operations. $\|\boldsymbol{x}\|_0$ accounts the nonzero number of $\boldsymbol{x}$ and $\|\boldsymbol{x}\|_2$ is the Euclidean norm of $\boldsymbol{x}$.

## II. SYSTEM MODEL AND PROBLEM FORMULATION

Assuming a $W/2$-bandwidth waveform $x(t)$ is transmitted over a frequency-selective fading channel h(t) with additive Gaussian noise $z(t)$, the received continuous-time signal waveform is obtained as

$$r(t) = \int_0^{\tau_{\max}} h(\tau)x(t-\tau)d\tau + z(t), \quad (1)$$

where $\tau_{\max}$ denotes the maximum time-delay spread of channel. According to the Shannon sampling theorem, channel comprises $N = \lceil W\tau_{\max} \rceil + 1$ sampling taps with the Nyquist rate sampling period $1/W$. Hence, the physical channel impulse response $h(t)$ can be approximated by $h(t) = \sum_{n=0}^{N-1} h_n \delta(t - n/W)$ [12]. Here, we assume that the N-length discrete channel vector $\boldsymbol{h} = [h_0, h_1, ..., h_{N-1}]^T$ is supported by only K dominant channel taps. Such a channel is often termed as K-sparse multipath channel ($K \ll N$). If we use high-speed ADC sampling receiver as shown in Fig. 2, then the $m$-th sampling coefficient $r_m$ is given by

$$r_m = \int_0^T r(t)f(m/W - t)dt, \ m = 1, 2, ..., M, \quad (2)$$

where $f(t) = \sin \pi t / \pi t$ denotes high-speed ADC sampling function and $T$ is the signal period. Hereby, the equivalent discrete-time system model is described by

$$\boldsymbol{r} = \boldsymbol{X}\boldsymbol{h} + \boldsymbol{z}, \quad (3)$$

where $\boldsymbol{r} = [r_1, r_2, ..., r_M]^T$ is an $M$-dimensional observed signal vector, $\boldsymbol{z}$ is an M-dimensional Gaussian noise samples of zero mean and variance $\sigma_n^2$, and $\boldsymbol{X}$ is a partial Toeplitz matrix of the form

$$\boldsymbol{X} = \begin{bmatrix} x_{N-1} & x_{N-2} & \cdots & x_1 & x_0 \\ x_N & x_{N-1} & \cdots & x_2 & x_1 \\ \vdots & \vdots & & \vdots & \vdots \\ x_{N+M-2} & x_{N+M-3} & \cdots & x_M & x_{M-1} \end{bmatrix} \quad (4)$$

Suppose that the receiver is equipped with $p$ low-speed ADCs as shown in Fig. 3. The integration period $T$ is then split into $P$ subintervals and $\boldsymbol{y}_m = [y_{m1}, y_{m2}, ..., y_{mP}]^T$, $m = 1, 2, ..., M$ denote the vectors of sub-sample collected against the sampling waveform, $\{x_m(t)\}_{m=1}^M$. The sub-sample coefficient $y_{mp}$ is then given by

$$y_{mp} = \int_{(p-1)T/P}^{pT/P} r(t)f(N/P \cdot (m/W) - t)dt, \quad (5)$$

where $m = 1, 2, ..., M$. Then the total number of subsamples collected by all parallel ADCs over all the subperiods is a $M \times P$ matrix which is shown in Fig. 4. These subsamples can be expressed as

$$\boldsymbol{Y} = \begin{bmatrix} y_{11} & y_{12} & \cdots & y_{1P} \\ y_{21} & y_{22} & \cdots & y_{2P} \\ \vdots & \vdots & \ddots & \vdots \\ y_{M1} & y_{M2} & \cdots & y_{MP} \end{bmatrix} \quad (6)$$

where the $m$-th row contains the subsamples obtained by correlating the measured signal with the waveform $x_m(t)$ over $P$ subperiods with $N/P$-length. Comparing with ADC based original M samples in Eq. (3), i.e., the sampling matrix $\boldsymbol{Y}$ collected at parallel low-speed ADCs over the whole signal duration $T$ in Eq. (6), the relationship between them is given as

$$r_m = \sum_{p=1}^{P} y_{mp}, \quad m=1,2,...,M. \quad (7)$$

In the sequence, we can easily find the following relation

$$y_{mp} = \sum_{(p-1)N/P}^{pN/P} x_{mn} h_n, \quad (8)$$

where $p=1,2,...,P$, $n=1,2,...,N$ and $m=1,2,...,M$. According to above Eq. (7-8), extra observation vector $\boldsymbol{r}_e = [r_{M+1}, r_{M+2},...,r_{M+M_e}]^T$ can be obtained, where $M_e$ is the extracted length and its $m$-th element $r_{M+m}$ is extracted from observation matrix, given by

$$\begin{aligned} r_{M+m} &= \sum_{p=1}^{P} y_{\{[(m+p-2) \bmod M]p\}} \\ &= \sum_{p=1}^{P} \sum_{(p-1)N/P}^{pN/P} x_{\{[(m+p-2) \bmod M]n\}} h_n, \end{aligned} \quad (9)$$

where 'mod' denotes modulo operation and $m=1,2,...,M_e$, $n=1,2,...,N$. According to the Eq. (7-9), we can obtain the following equation

$$\boldsymbol{\phi} = \boldsymbol{\Phi} \boldsymbol{h} + \boldsymbol{\gamma}, \quad (10)$$

where $\boldsymbol{\phi} = [\boldsymbol{r}^T, \boldsymbol{r}_e^T]^T$ denotes equivalent received signal vector, $\boldsymbol{\Phi} = [\boldsymbol{X}^T, \boldsymbol{X}_e^T]^T$ denotes overall training matrix, and $\boldsymbol{\gamma} = [\boldsymbol{z}^T, \boldsymbol{z}_e^T]^T$ is additive noise.

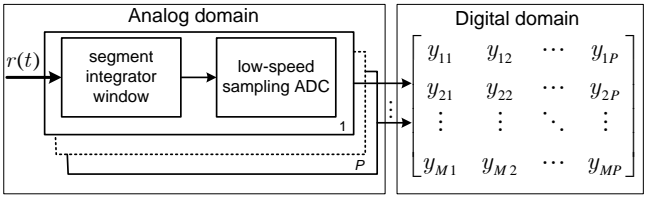

Fig. 4. Low-speed ADC working at sub-Nyquist rate sampling.

### III. HIGH-RESOLUTION COMPRESSIVE CHANNEL ESTIAMTION

We formulate the sub-Nyquist rate sampling based sparse channel estimation as a compressive sensing problem [8,9]. Sparse channel estimation methods have been intensively studied in recent years [11]. To make comparison with our previously proposed method, sparse channel estimation is implemented by CoSaMP algorithm. The detail of our proposed method is introduced as follows:

Given the received signal vector $\boldsymbol{\phi}$, equivalent training matrix $\boldsymbol{\Phi}$, the number of dominant channel taps is set to $K$. The proposed method is composed of four steps:

*Initialization*. Set the dominant taps index set $\Omega_0 = \varnothing$, the residual estimation error $\boldsymbol{r}_0 = \varnothing$ and put the initialize iteration counter as $i = 1$

*Identification*. Select a column subset $\Omega_i$ of $\boldsymbol{\Phi}$ that is most correlated with the residual:

$$\Omega_i = \arg\max \left| \langle \boldsymbol{r}_{i-1}, \boldsymbol{\Phi} \rangle \right|, \text{ and } \Omega_i = \Omega_{i-1} \cup \Omega_i. \quad (11)$$

Using LS method to calculate a channel estimator as $\Omega_{LS} = \arg\min \|\boldsymbol{\phi} - \boldsymbol{\Phi} \boldsymbol{h}\|_2$, and select $K$ maximum dominant taps denoted by $\boldsymbol{h}_{LS}$. The positions of the selected dominant taps are denoted by $\Omega_{LS}$.

*Merging*. The positions of dominant taps are merged by $\Omega_i = \Omega_{LS} \cup \Omega_i$.

*Estimation*. Compute the best coefficient to approximate the channel vector with chosen columns,

$$\hat{\boldsymbol{h}}_i = \arg\min_{\hat{\boldsymbol{h}}} \left\| \boldsymbol{\phi} - \boldsymbol{\Phi}_{\Omega_i} \boldsymbol{h} \right\|_2. \quad (12)$$

*Pruning*. Select the $\Omega_i$ largest channel taps of $\boldsymbol{h}_i$ and set

$$\hat{\boldsymbol{h}}_{\Omega \setminus \Omega_i} = 0. \quad (13)$$

*Iteration*. Update the estimation error:

$$\boldsymbol{\phi}_i = \boldsymbol{\phi} - \boldsymbol{\Phi}_{\Omega_i} \hat{\boldsymbol{h}}_i, \quad (14)$$

increase the iteration counter $i$. Repeat (11-14) until stopping criterion is satisfied and then set $\hat{\boldsymbol{h}} = \boldsymbol{h}_i$.

### IV. NUMERICAL SIMULATIONS

In this section, we will compare the performance of the proposed estimators with 10000 independent Monte-Carlo runs for averaging. The length of sparse multipath channel $\boldsymbol{h}$ is set as $N = 96$ and its number of dominant taps is set as $K$. We consider two kinds of distributions on all dominant channel taps that their values are generated from $1/K$-uniform distribution and random Gaussian distribution. The positions of dominant channel taps are randomly allocated within the length of $\boldsymbol{h}$ and is subjected to $E[\|\boldsymbol{h}\|_2^2] = 1$. The initial $\boldsymbol{X}$ is an equivalent $M \times N$ partial Toeplitz matrix. The initial length of training sequence is set as $M = 32$ and the extract training length is $M_e$. The number of parallel low-speed ADCs is set as $P = 8$. The received SNR is defined as $10\log(E_0/\sigma_n^2)$, where $E_0$ is received power. Here, we set the SNR values as 10dB, 20dB and 30dB in the following numerical simulations.

The estimation performance is evaluated by two criterions: successful recovery of dominant channel taps and average mean square error (Average MSE). Average MSE of channel estimators $\hat{\boldsymbol{h}}$ is defined by

$$Average \ MSE(\hat{\boldsymbol{h}}) = \frac{E\left\{\|\boldsymbol{h} - \hat{\boldsymbol{h}}\|_2^2\right\}}{L}, \quad (15)$$

where $\boldsymbol{h}$ and $\hat{\boldsymbol{h}}$ denote the original channel vector and its estimator, respectively. In the numerical results, traditional method and lower bound based on high rate sampling ADC, utilizing different equivalent random partial Toeplitz matrices $\boldsymbol{X}$ with training length $M$ (short training sequence) and $M + M_e$ (long training sequence) are calculated, respectively. To make proposed method uses the same training matrix $\boldsymbol{X}$ with length $M$ (short training sequence) and its extracted training matrix $\boldsymbol{X}_e$.

In Fig. 5 and Fig. 6, we compare the successful recovery probability of dominant channel taps as a function of extracted training signal length from 0 to 56. The received SNR is set as 10dB, 15dB and 20dB, respectively. Here, the number of dominant channel taps is set as $K = 4$. From the two figures, it is observed that our proposed method is very close to optimal performance bound and the successful recovery increases as the extracted training length increases, while the traditional method is invariant and the performance is very poor. In addition, the two figures also showed that that channel taps are related with their distribution, e.g., uniform distributed channel taps are reconstructed easier than random distributed Gaussian taps.

Furthermore, the average MSE of different methods are also depicted in Fig. 7 and Fig. 8. From the two figures, it is observed that the estimation performance of our proposed method is better than traditional method and close to the lower bound. According to the above numerical simulations, the effectiveness of sub-Nyquist sampling rate ADC based method has been verified.

In the next, the relationship between the proposed method and channel sparsity is considered. Since the space limitation, here, we only consider the uniform distribution of dominant taps on compressive channel estimation. Assume that the number of channel length is same, while the number of dominant taps is 2, 4, 6 and 8, respectively. We also compare their recovery probability of dominant channel taps (see Fig. 9) and average MSE (see Fig. 10). From the two figures, whatever the channel sparsity, our proposed method can always close to their lower bound. Hence, the proposed method is stable for the sparse channel with different number of dominant channel taps.

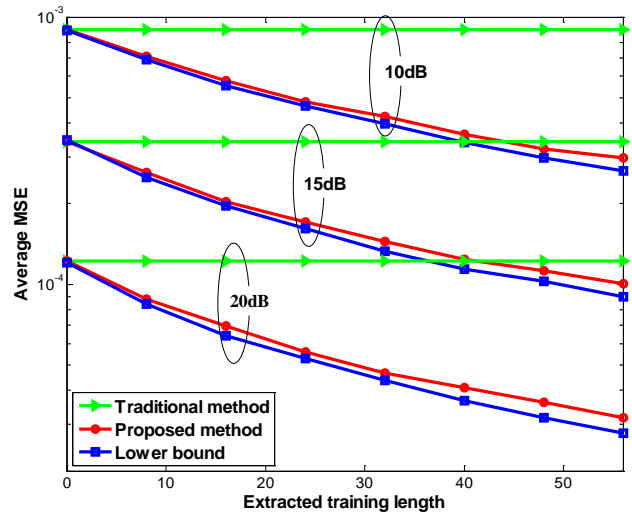

Fig.7. Average MSE performance versus the extracted training length, where the dominant channel taps satisfy 1/K-uniform distribution.

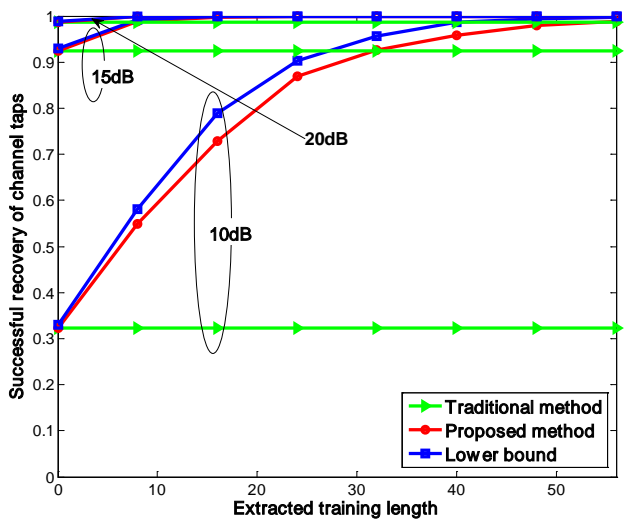

Fig.5. Successful recovery probability of the dominant channel taps versus the extracted training length, where the dominant channel taps satisfy uniform distribution.

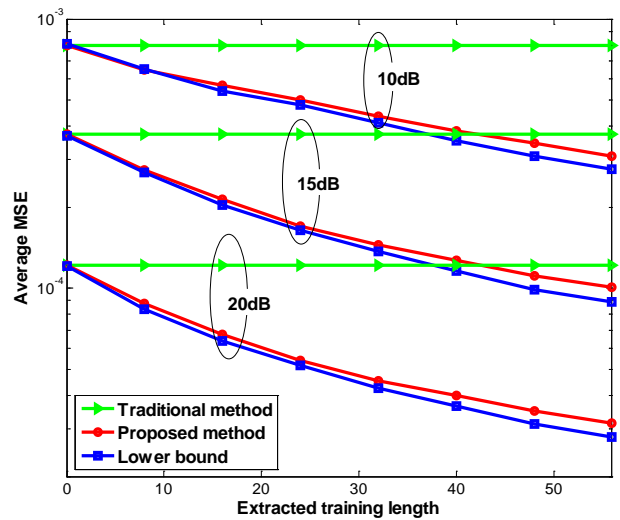

Fig.8. Average MSE performance versus the extracted training length, where the dominant channel taps satisfy Gaussian distribution.

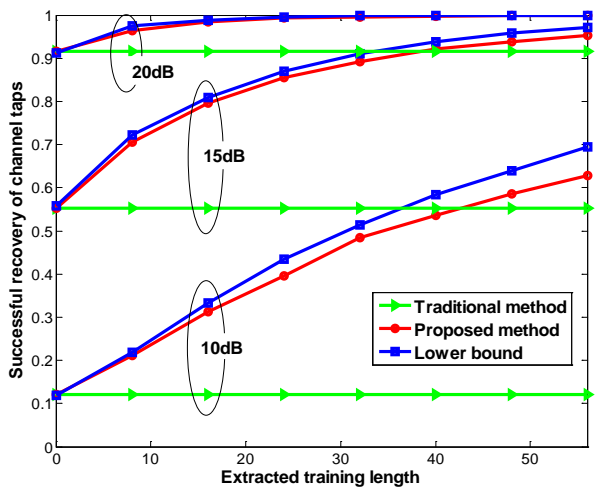

Fig.6. Successful recovery probability of the dominant channel taps versus the extracted training length, where the dominant channel taps satisfy Gaussian distribution

## V. CONCLUSION

In this paper, we have investigated high-resolution compressive channel estimation based on the sub-Nyquist rate sampling ADC. First of all, we formulated the channel estimation as compressive sensing problem. In the sequence, an high-resolution compressive channel estimation method has been proposed for sparse multipath broadband communication systems. Comparison with the traditional sparse channel estimation methods has shown that, our proposed method has two advantages: 1) can achieve high-resolution channel estimation and even much better estimation performance; 2) can save communication cost. In the future work, we will study on multi-antenna high-resolution compressive channel estimation based on sub-Nyquist sampling ADC.

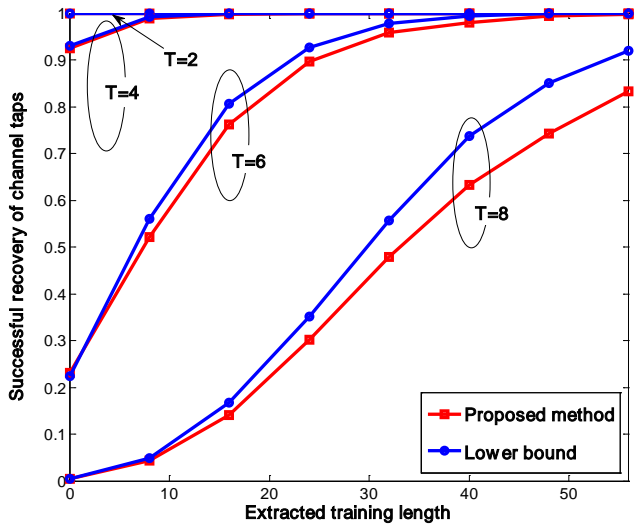

Fig.9. Average MSE performance versus the extracted training length, where the dominant channel taps satisfy uniform distribution.

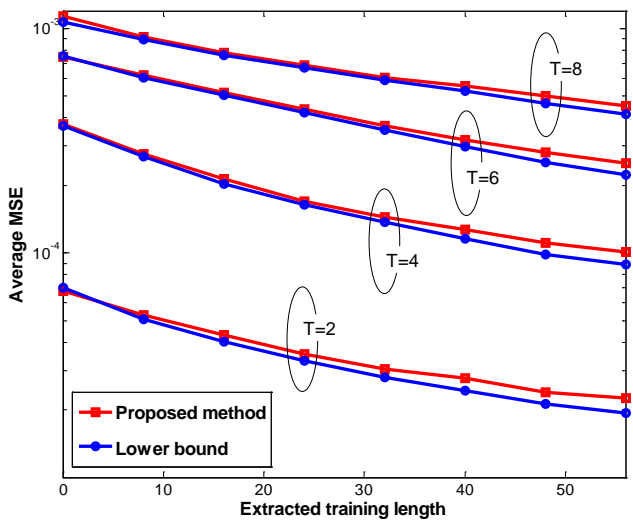

Fig. 10. Average MSE performance versus the extracted training length, where the dominant channel taps satisfy uniform distribution.

ACKNOWLEDGMENT

The authors would like to thank Prof. Guan Yong Liang for helpful discussions. This work is supported in part by ``the Fundamental Research Funds for the Central Universities'' under award number ZYGX2010J016.